\documentclass[a4paper,preprintnumbers,floatfix,superscriptaddress,pra,twocolumn,showpacs,notitlepage]{revtex4}
\usepackage[utf8]{inputenc}
\usepackage{graphicx}
\usepackage{amsmath,amsthm,mathpazo}
\usepackage{array}
\usepackage{xcolor}
\newtheorem{theorem}{Theorem}
\newtheorem{result}[theorem]{Result}

\begin{document}
\title{Information Causality without concatenation}
\author{Nikolai Miklin}
\affiliation{International Centre for Theory of Quantum Technologies (ICTQT), University of Gdansk, 80-308 Gda\'nsk, Poland}
\author{Marcin Paw{\l}owski}\thanks{marcin.pawlowski@ug.edu.pl}
\affiliation{International Centre for Theory of Quantum Technologies (ICTQT), University of Gdansk, 80-308 Gda\'nsk, Poland}

\begin{abstract}
    Information Causality is a physical principle which states that the amount of randomly accessible data over a classical communication channel cannot exceed its capacity, even if the sender and the receiver have access to a source of nonlocal correlations. This principle can be used to bound the nonlocality of quantum mechanics without resorting to its full formalism, with a notable example of reproducing the Tsirelson's bound of the Clauser-Horne-Shimony-Holt inequality. Despite being promising, the latter result found little generalization to other Bell inequalities because of the limitations imposed by the process of concatenation, in which several nonsignaling resources are put together to produce tighter bounds. In this work, we show that concatenation can be successfully replaced by limits on the communication channel capacity. It allows us to re-derive and, in some cases, significantly improve all the previously known results in a simpler manner and apply the Information Causality principle to previously unapproachable Bell scenarios.
\end{abstract}

\maketitle

\section{Introduction}

Information Causality (IC) is a physical principle proposed to bound nonlocality of correlations without resorting to the full formalism of quantum mechanics~\cite{IC,ICrev}. Instead, the bounds are derived only from the axioms of information theory. In a nutshell, the principle states that if one party has a single use of a communication channel with a capacity $C$ to send the other party some information, then the amount of information potentially available to the receiver cannot exceed $C$ even if the parties share some nonlocal resources. In Ref.~\cite{IC} it was shown that both classical and quantum information theories, and every generalization of them adhering to some of their intuitive properties, obey IC. At the same time, the principle of IC is strong enough to partially recover the boundary of the set of quantum nonlocal correlations~\cite{allcock2009recovering}. Most notably, as shown in Ref.~\cite{IC}, IC can be used to re-derive the Tsirelson's bound of the Clauser-Horne-Shimony-Holt (CHSH) inequality, answering the long-standing question of Popescu and Rorhlich on the reason for its value~\cite{Tsirelson,CHSH,PR}. Moreover, IC was shown to rule out stronger-than-quantum correlations, which could not be detected by other bipartite principles, such as Macroscopic Locality~\cite{ML,Bassard2006,Valerios}. Finally, the principle of IC is the only known candidate with the potential to recover the exact boundary of the set of bipartite nonlocal quantum correlations~\cite{AlmostQ}.

The problem that one faces when deriving bound on the strength of nonlocal correlations from IC is that one has to find a suitable communication protocol that makes use of those correlations. Until now, it was believed that to obtain the strongest results, one must use protocols that allow for concatenation, the process of combining the outcomes of several copies of the nonlocal source in a way that increases the amount of potentially available information in stronger-than-quantum theories. This requirement significantly limits the types of Bell scenarios for which the bounds from IC can be derived~\cite{IC,EARAC,Valerios}.  

In this work, we argue that the procedure of concatenation is not required and often suboptimal in proofs utilizing IC. More precisely, we show that by considering a non-identity communication channel with suitably chosen capacity and a single copy of a nonlocal resource, one can: (a) re-derive all the results from Ref.~\cite{IC,EARAC}; (b) tighten the bounds found in Ref.~\cite{Valerios}; (c) apply IC to Bell scenarios for which no concatenation procedure is known. We expect that with the modified construction, IC is likely to become a staple tool for finding bounds on Bell nonlocality in situations when traditionally applied numerical methods are computationally demanding~\cite{NPA}. 

\section{IC and concatenation}
We start by restating the formulation of the IC principle from Ref.~\cite{ICrev}. Consider a communication scenario in which the sender has $N$ real-valued random variables $a_0,...,a_{N-1}$ and a single use of a channel with classical communication capacity $C$. Then 
\begin{equation} \label{basicIC}
    \sum_{i=0}^{N-1} I(a_i;b_i)\leq C,
\end{equation}
where $b_i$ is a random variable denoting the receiver's guess of the value of $a_i$ if the receiver chooses to guess it, and $I(a_i;b_i)$ is the Shannon's mutual information.

In the original formulation of IC, as given by Ref.~\cite{IC}, the bound on the right-hand side of Eq.~(\ref{basicIC}) is defined as the size of a message that the sender communicates in each round. The latter, somewhat vague statement, was later clarified in authors' subsequent work of Ref.~\cite{ICrev}: {\it ``Notice that it does not matter how this information is encoded: when we refer to `sending the $M$ bit message', it should be understood as a single use of a channel with classical communication capacity $M$.''}

In order to apply IC to quantum correlations, in Ref.~\cite{IC} the authors start by considering the following protocol~\cite{vanDam}: The parties share a pair of devices characterized by probability distribution $P(a,b|x,y)$ with all variables $a,b,x,y$ binary, $x$ and $a$ being the input and output of the sender, and $y$ and $b$ -- the input and output of the receiver (Here and later in the text, by $P(a,b|x,y)$ we mean all the probabilities $P(a=i,b=j|x=k,y=l)$,$\forall i,j,k,l$). Let $N=2$ and $P(a,b|x,y)$ be such that $b=a\oplus x\cdot y$ with probability $p$ for both $y\in\{0,1\}$ ($\oplus$ denotes the summation modulo $2$). The sender chooses $x=a_0\oplus a_1$ and transmits a message $m=a_0\oplus a$ to the receiver. In order to learn about $i$-th bit $a_i$, the receiver chooses $y=i$ and computes $b_i=m\oplus b$. It is straightforward to confirm that $a_i=b_i$ with probability $p$. If values of $a_0$ and $a_1$ are distributed uniformly, this protocol yields $I(a_i;b_i)=1-h(p)$, where $h(.)$ is the Shannon's binary entropy. 
If $m$ is send over a perfect channel, then $C=1$ since $m$ is just a bit. The bound resulting from Eq.~(\ref{basicIC}) is $2(1-h(p))\leq 1$, which implies $p\leq 0.890$. It is easy to see that the CHSH expression for $P(a,b|x,y)$ is also equal to $p$~\cite{CHSH}. Hence, we have derived a nontrivial bound on CHSH inequality from IC. However, the maximum quantum value of $p$, known as the Tsirelson bound, is $p_Q=\frac{1}{2}\left(1+\frac{1}{\sqrt{2}} \right)\approx 0.854$~\cite{Tsirelson}, which is significantly lower than what we have just derived.

To obtain a tighter bound on $p_Q$ from IC, the authors of Ref.~\cite{IC} propose to increase $N$, the number of bits given to the sender at each round, and use concatenation. Concatenation is essentially a process of locally combining inputs and outputs of many copies of the same pair of devices. It is, however, different from the simple ``wiring'' of devices~\cite{Wiring}, as in the case of concatenation the input of each device of the sender also depends on the bits $a_i$.  

Here, we do not give the explicit description of concatenation, and refer the reader to Ref.~\cite{IC}. Instead, we only state some details of it as facts. This procedure works for $N=2^k$ for some positive integer $k$. At sender's the devices are placed in layers with $k$-th layer having $2^{k-1}$ devices. Each of the devices produces a ``message'' just like in the protocol above and the resulting $2^{k-1}$ ``messages'' are taken as inputs for the devices in layer $k-1$. Each layer of concatenation introduces an error diluting the information about each of the bits $a_i$. If at some layer $k-1$ the probability of successfully decoding $a_i$ is $p_{k-1}$, at the next level $k$ it will be \begin{equation} \label{next}
p_{k}=p_{k-1} p+(1-p_{k-1})(1-p)=\frac{1+e_{k-1} e}{2},    
\end{equation}
where $e_{k-1}$ and $e$ are biases of $p_{k-1}$ and $p$, i.e., $p_{k-1}=\frac{1+e_{k-1}}{2}$ and $p=\frac{1+e}{2}$. The condition in Eq.~(\ref{basicIC}) for a protocol with $k$ levels of concatenation becomes
\begin{equation}
    2^k\left(1-h\left(\frac{1+e^k}{2} \right) \right)\leq 1.
\end{equation}
For every $k$ it puts a lower bound on $e$ and for $k\to \infty$ the bound on $e$ reaches $\frac{1}{\sqrt{2}}$ while the bound on $p$ converges to our aim, $p_Q$.

\section{Replacing concatenation with noisy channel}
\begin{figure}
    \centering
    \includegraphics[width=8cm]{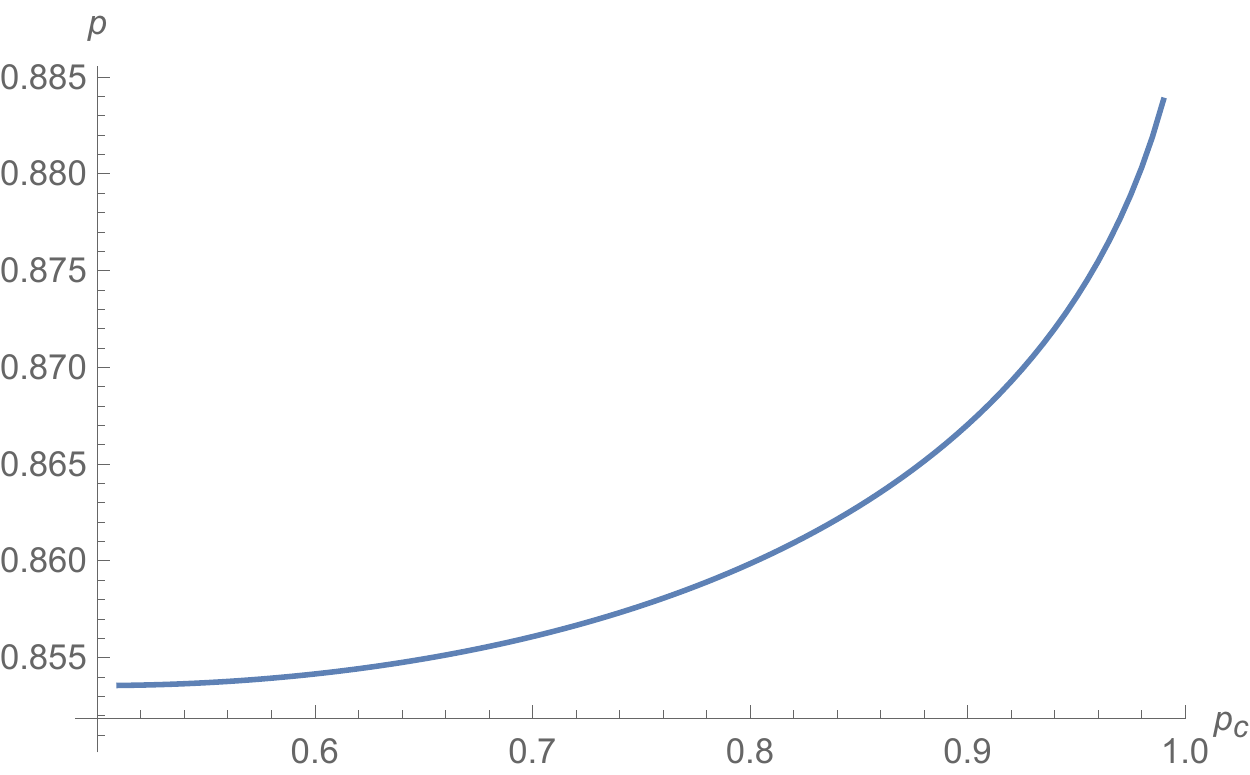}
    \caption{Bound on $p$ as a function of $p_c$ characterizing binary symmetric channel. $p\to p_Q$ as $p_c\to \frac{1}{2}$.}
    \label{fig:1}
\end{figure}

There is, however, an easier way to obtain the bound $p_Q$ from IC without the need of concatenation. Let us start by considering the same protocol as before for a single pair of devices, but change the communications channel between the sender and the receiver to a binary symmetric one. Such a channel transmits an unchanged input bit with probability $p_c$, and with probability $1-p_c$ it returns the flipped bit. The capacity of this channel is $1-h(p_c)$. The probability for $b_i$ to be equal to $a_i$ is obtained with the same formula in Eq.~(\ref{next}), where instead of $p_{k-1}$ we use $p_c$, yielding $\frac{1+e_c e}{2}$. The condition in Eq.~(\ref{basicIC}), expressed in the terms of biases becomes
\begin{equation} \label{bsc}
    2\left(1-h\left(\frac{1+e_c e}{2} \right) \right)\leq 1-h\left(\frac{1+e_c}{2} \right). 
\end{equation}
The bound implied by the above condition becomes stronger as $p_c$ and the channel capacity decrease as shown in Fig.~\ref{fig:1}. For $p_c$ approaching $\frac{1}{2}$ the bound on $p$ approaches $p_Q$. As we show below, this is a consequence of a more general result.

\begin{result}\label{res1}
 Any bound on nonlocality in the case of unbiased errors obtained with concatenation procedure can also be obtained by a protocol involving a single pair of devices and a suitably chosen discrete memoryless channel.
\end{result}
Before we move to the proof, we need to clarify what we mean by ``unbiased errors case''. Let us consider a single pair of the devices producing probability distribution $P(a,b|x,y)$, where $a,b\in\{0,1,\dots,d-1\}$, and $y\in\{0,1,\dots,n-1\}$. Defining the range of $x$ is not necessary for this argument. 
Let us assume a protocol involving the communication of one of $d$ possible messages over a classical identity channel such that with probability $p$, the receiver makes the right guess $b_i=a_i$.  The ``unbiased errors'' assumption means that $p$ does not depend on $i\in\{0,1,\dots,n-1\}$, every term in $p$ corresponding to each value of $a_i$ is equal, and that all the other ``error'' cases $b_i\neq a_i$ are equally probable and also uniformly distributed with respect to $a_i$. Arguably, the case of unbiased errors is a special one, but it is general enough to encompass all currently known results~\cite{IC,EARAC,Valerios}. 

Let us now clarify what we mean by finding ``bound on nonlocality''. Given $P(a,b|x,y)$, one may ask whether this nonlocal behavior complies with IC's statement, giving a ``yes/no" answer. However, one may instead ask a quantitative question of how much noise needs to be added to $P(a,b|x,y)$ in order for it to satisfy IC. It is standard to consider the white noise, and the guessing probability $p$ in IC is proportional to the amount of white noise required. Hence, $p$ quantifies nonlocality of $P(a,b|x,y)$. In some cases, as it is for CHSH, the bound on $p$ also implies the bound on Bell inequality.

\begin{proof}
Following Ref.~\cite{Valerios} we generalize the definition of the bias $e$ of probability $p$ as 
\begin{equation} \label{gen:bias}
   p=\frac{1+(d-1)e}{d}. 
\end{equation}
If we choose to concatenate the protocol $k$ times, as it is done in Refs.~\cite{EARAC,Valerios}, the success probability of $b_i=a_i$ will be $p_k=\frac{1+(d-1)e^k}{d}$. From the symmetry of the protocol and unbiasedness of errors, we conclude that all the mutual information terms in the IC expression are equal, and given by $I(a_i;b_i) = I_d(e^k)$, where 
\begin{equation} \label{Id}
\begin{split}
    I_d(e)&=\log d-h\left(\frac{1+(d-1)e}{d}\right)\\
    &-\frac{(d-1)(1-e)}{d}\log (d-1).
    \end{split}
\end{equation}
The above expression is known as Fano's inequality~\cite{fano}, which is equality in this case (See Appendix~\ref{app:fano} for a short proof).
Let us assume that $k$ is such that $k$ levels of concatenation are not enough to demonstrate that $p$ violates IC, but $k+1$ are. In other words:
\begin{equation}
\begin{split}
       n^kI_d(e^k) \leq \log d \\
    n^{k+1}I_d(e^{k+1})> \log d,
\end{split}
\end{equation}
which implies
\begin{equation} \label{cond:conc}
n I_d(e^{k+1})>I_d(e^k).    
\end{equation}

Let us now take a single pair of the devices and let the parties communicate over a discrete memoryless channel channel with uniform errors, i.e., with probability $p_c$ the message is unchanged and with $1-p_c$ probability it is changed to one of the other $d-1$ messages in accordance with the uniform distribution. Let $p_c=\frac{1+(d-1)e^k}{d}$, then the capacity of the channel is $C=I_d(e^k)$ (See Appendix~\ref{app:fano} for a short proof). The probability for $a_i=b_i$ is $p_c p+\frac{(1-p_c)(1-p)}{d-1}=\frac{1+(d-1)e^{k+1}}{d}$ and $I(a_i;b_i)=I_d(e^{k+1})$, for all $i\in\{0,1,\dots,n-1\}$. Therefore, in this case the IC condition in Eq.~(\ref{basicIC}) reads
\begin{equation}
    n I_d(e^{k+1})\leq C= I_d(e^k).
\end{equation}
If the probability $p$ is such that Eq.~(\ref{cond:conc}) holds, the above inequality will be violated, which means that $p$ can also be detected by the IC with a single pair of the devices and a discrete memoryless channel with capacity $C$.
\end{proof}

\section{New and tighter bounds}
In this section, we demonstrate that the bounds on nonlocality obtained with the new approach are strictly better in some cases. We also discuss cases in which this approach can provide bounds while the concatenation procedure is not applicable.

In the proof of Result~\ref{res1}, instead of choosing $p_c$ to be the success probability corresponding to $k$-th level of concatenation, we can take $p_c = \frac{1+(d-1)e_c}{d}$ and optimize over $e_c$. The condition that we use to bound nonlocality is the following
\begin{equation} \label{gsc}
    n I_d(e_ce)\leq I_d(e_c).
\end{equation}
Even though it is cumbersome to write the solution for the optimal $e_c$ explicitly, the optimization over a single real parameter can be done up to an arbitrary numerical precision. In table~\ref{tb} we compare the bounds on the bias $e$ of the success probability $p$ implied by Eq.~(\ref{gsc}) with the bounds from Ref.~\cite{Valerios}, calculated using concatenation procedure for $n=2$. 

\begin{table}
\begin{tabular}{ | m{1cm} | m{2cm}| m{1cm} | m{1cm} | } 
\hline
 & Optimal $e_c$ & $e$ & $e'$ \\ 
\hline
d=3 & 0.295 & 0.702 & 0.708  \\ 
\hline
d=4 & 0.389 & 0.696 & 0.705 \\ 
\hline
d=5 & 0.436 & 0.690 & 0.700 \\ 
\hline
d=20 & 0.531 & 0.648 & 0.659 \\ 
\hline
\end{tabular}
\caption{Optimal $e_c$ is the value for which Eq.~(\ref{gsc}) puts the tightest bound on $e$. $e$ is the improved bound, while $e'$ is the bound from Ref.~\cite{Valerios}.}
\label{tb}
\end{table}

The major challenge of finding bounds on nonlocality with the concatenation procedure is calculating each level's success probabilities. This calculation is easy only if one assumes the unbiasedness of errors, as described in Result~\ref{res1}. This assumption puts a significant constraint on the choice of the protocol and the correlations $P(a,b|x,y)$ that can be easily bounded using the IC. 

On the contrary, the method suggested in this paper can be applied to any protocol. Below we give an example of bounding nonlocality in a Bell scenario with $3$ settings per party and with binary outcomes (often referred to as $3322$ scenario). Let us consider a non-signaling distribution $P_{NS}(a,b|x,y)$ ($x,y\in\{0,1,2\}$) given by the following relation
\begin{equation}\label{eq:box_3322}
    a\oplus b = 
    \begin{cases}
    1,& \text{if } (x,y)\in\{(1,2),(2,1),(2,2)\}\}\\
    0,              & \text{otherwise},
\end{cases}
\end{equation}
with all marginal probability distribution being uniform, i.e., $\sum_{i}P(a=i,b=j|x,y) = \sum_jP(a=i,b=j|x,y) = \frac{1}{2}, \forall i,j$. This distribution gives the maximal violation equal to $1$ of the $I_{3322}$ inequality~\cite{Collins_2004}. Let us now consider a pair of devices producing correlations $P_e(a,b|x,y) = eP_{NS}(a,b|x,y)+(1-e)P_L(a,b|x,y)$, where $P_L(a,b|x,y)$ is the white noise distribution for which $P_L(a=i,b=j|x,y)=\frac{1}{4}, \forall i,j$ and all values of $x,y$. The value of $I_{3322}$ inequality for the considered mixing is $2e-1$. We ask a question of the maximal degree of nonlocality, specified by $e$, that is allowed by IC. We can find a protocol, which we specify in Appendix~\ref{app:3322}, that is optimal for the nonlocal correlations $P_{NS}(a,b|x,y)$ from Eq.~(\ref{eq:box_3322}). Using the symmetric channel, parameterized by $e_c$, in the limit of $e_c\rightarrow 0$ we obtain a bound $e\rightarrow\frac{2}{3}$, which is not far from the quantum bound of $\frac{3}{5}$, which can be confirmed (up to a numerical precision) by the hierarchy of semidefinite programming of Ref.~\cite{NPA}. To compare, the bound which can be derived from IC with a channel capacity of 1 is about $0.7445$, and there is no clear way to construct a concatenation procedure for this protocol or to calculate the corresponding guessing probabilities.

\section{Discussion}
We have shown that concatenation can be successfully replaced by considering different classical communication channels in the protocols used to bound nonlocality with IC. Apart from showing that all the results already obtained can be re-derived with the new approach, we also showed that some could be improved. Additionally, we gave an example of a scenario that would be very challenging to approach with the concatenation procedure. Perhaps the most important goal of this paper is to show that IC's scope of applicability can be significantly widened. Therefore, our paper opens many possible ways for future work. 

We list some of the open questions which we believe deserve a separate study. In the current paper, we limited ourselves to a particular type of discrete memoryless channels, namely symmetric channels. These channels seem very well suited to analyze Bell inequalities, which correspond to unique games, that is games in which for every combination of parties' inputs and one party output, there is a unique output for the other party that wins the game. However, for other Bell inequalities, different channels can yield stronger results, and find optimal ones is an open question. Additionally, considering random data in the IC scenario with different alphabets could be used to bound nonlocality in distributions with a bias towards one of the outcomes. Finally, the statement of IC can be read in the opposite direction. Namely, given a value of guessing probability, one can obtain a lower bound on the minimal communication required for such correlations, which can be used for randomness certification.

\section{Acknowledgements}
We acknowledge support by the Foundation for Polish Science (IRAP project, ICTQT, contract no. 2018/MAB/5, co-financed by EU within Smart Growth Operational Programme), M.P.~acknowledges the support of NCN through grant SHENG (2018/30/Q/ST2/00625). This research was made possible by funding from QuantERA, an ERA-Net cofund in Quantum Technologies (www.quantera.eu), under the project eDICT.

\begin{appendix}
\section{Technical details of the proof of Result~\ref{res1}}
\label{app:fano}
Fano's inequality for two random variables $a$ and $b$ taking values in $\{0,1,\dots,d-1\}$ can be written as follows
\begin{equation} \label{eq:fano}
    I(a;b)\geq\log d-h\left(p\right)-(1-p)\log (d-1),
\end{equation}
where $p = \mathrm{P}(a=b)$. In Eq.~(\ref{Id}) we took $p=\frac{1+(d-1)e}{d}$ and denoted the function on the right-hand side of Eq.~(\ref{eq:fano}) as $I_d(e)$. Let $P(a=i,b=j) = r(j|i)P(a=i)$ be the decomposition of the joint probability distribution of $a$ and $b$ with a response function (conditional distribution) $r(j|i)$. For the case of probabilities in the derivations of Result~\ref{res1}, the form of $r(j|i)$ is the following: $r(i|i) = p,\forall i$, and $r(j|i) = \frac{1-p}{d-1}, \forall j\neq i$. The mutual information $I(a;b)$ by definition is equal to
\begin{equation}
    I(a;b) = \sum_{i=0}^{d-1}\sum_{j=0}^{d-1}P(a=i)r(j|i)\log\frac{r(j|i)}{\sum_{i=0}^{d-1}r(j|i)P(a=i)},
\end{equation}
Since the input $a$ is always taken to be uniformly distributed, we have that $P(a=i) = \frac{1}{d}$, and hence $\sum_{i=0}^{d-1}r(j|i)P(a=i) = \frac{1}{d},\forall j$. From here we can deduce that
\begin{equation}
\begin{split}
        I(a;b) &= \log(d)+\frac{1}{d}\sum_{i=0}^{d-1}\sum_{j=0}^{d-1}r(j|i)\log(r(j|i)) \\
        &=\log(d)+p\log(p)+(1-p)\log\left(\frac{1-p}{d-1}\right),
\end{split}
\end{equation}
which is exactly equal to the right-hand side of Eq.~(\ref{eq:fano}).

Now, we give a short proof that capacity of discrete memoryless channel, which transmits $d$-dimensional message unchanged with probability $p_c = \frac{1+(d-1)e_c}{d}$ and with probability $1-p_c$ changes it to one of the $d-1$ other values according to the uniform distribution, is equal to $I_d(e)$ (given by the right-hand side of Eq.~(\ref{eq:fano})). 

By definition, the capacity of a discrete memoryless channel is
\begin{equation}
    C = \max_{P(a)}I(a;b),
\end{equation}
where $b$ is a guess of $a$, and $P(a)$ is the distribution of $a$. The proof is similar to the one above, since the response function $r(j|i)$ is exactly the same for the considered channel (where we substitute $p$ with $p_c$). The only part which requires the proof is that the optimal distribution $P(a)$ is the uniform one. It can be easily seen from the fact that, first of all, entropy is a concave function, and that the form of $r(j|i)$ is symmetric.

\section{Protocol for the 3322 scenario}
\label{app:3322}
Here, we give a specification of a protocol for the $3322$ scenario. In this protocol, the sender has access to three bits $a_0,a_1,a_2$, depending on which the choice of measurement $x\in\{0,1,2\}$ is determined. The bit message $m$ is a function of $a_0,a_1,a_2$ and the sender's outcome $a$. The decoding function determines the guess $b_i$ of $a_i$, depending on $i\in\{0,1,2\}$. The receiver chooses the measurement according to $i$, in the most obvious way $y=i$. The decoding functions are $b_0=m\oplus b\oplus 1$, $b_1=m\oplus b$, and $b_2=m\oplus b\oplus 1$. The message $m=a_0\oplus a\oplus 1$.
Below we specify the function for $x$ by a truth table.
\begin{equation}
    \begin{tabular}{c c c|c}
         $a_0$ & $a_1$ & $a_2$ & $x$\\ \hline
         $0$ & $0$ & $0$ & $0$\\
         $0$ & $0$ & $1$ & $2$\\
         $0$ & $1$ & $0$ & $0$\\
         $0$ & $1$ & $1$ & $1$\\
         $1$ & $0$ & $0$ & $1$\\
         $1$ & $0$ & $1$ & $0$\\
         $1$ & $1$ & $0$ & $2$\\
         $1$ & $1$ & $1$ & $0$.
    \end{tabular}
\end{equation}
The above protocol was obtained using simulated annealing~\cite{khachaturyan1979statistical}.
\end{appendix}

\bibliography{ic}
\end{document}